# Exploring the Efficacy of Partial Denoising Using Bit Plane Slicing for Enhanced Fracture Identification: A Comparative Study of Deep Learning-Based Approaches and Handcrafted Feature Extraction Techniques

Snigdha Paul
Electronics and Communication Engineering
Heritage Institute of Technology
Kolkata, India
snigdha.paul.ece25@heritageit.edu.in

Sambit Mallick
Electronics and Communication Engineering
Heritage Institute of Technology
Kolkata, India
sambit.mallick.ece25@heritageit.edu.in

Anindya Sen
Electronics and Communication Engineering
Heritage Institute of Technology
Kolkata, India
anindya.sen@heritageit.edu

*Abstract*— Computer vision has transformed medical diagnosis, treatment, and research through advanced image processing and machine learning techniques. Fracture classification, a critical area in healthcare, has greatly benefited from these advancements, yet accurate detection is challenged by complex patterns and image noise. Bit plane slicing enhances medical images by reducing noise interference and extracting informative features. This research explores partial denoising techniques to provide practical solutions for improved fracture analysis, ultimately enhancing patient care. The study explores deep learning model DenseNet, and handcrafted feature extraction. Decision Tree and Random Forest, were employed to train and evaluate distinct image representations. These include the original image, the concatenation of the four bit planes from the LSB, the four bit planes from the MSB, the fully denoised image, and an image consisting of six bit planes from MSB and two denoised bit planes from LSB. The purpose of forming these diverse image representations is to analyze SNR as well as classification accuracy and identify the bit planes that contain the most informative features. Moreover, the study delves into the significance of partial denoising techniques in preserving crucial features, leading to improvements in classification results. Notably, this study shows that employing the Random Forest classifier, the partially denoised image representation exhibited a testing accuracy of 95.61%, surpassing the performance of other image representations. These numerical results underscore the effectiveness of the proposed method in accurately identifying fractures. The outcomes of this research provide valuable insights into the development of efficient preprocessing, feature extraction and classification approaches for fracture identification. By enhancing diagnostic accuracy, these advancements hold the potential to positively impact patient care and overall medical outcomes.

## I. INTRODUCTION

### A. Motivation

In recent years, computer vision has transformed medical imaging, enabling automated analysis and interpretation of medical images for enhanced diagnosis and treatment. Fracture classification is a critical area where computer vision excels but faces challenges due to complex fracture patterns and image noise. To address the challenges of enhancing fracture images and extracting pertinent information, image processing techniques have emerged as potent tools. While computer vision has shown promising results in various aspects of medical imaging, the application of bit plane slicing and partial denoising methods for fracture classification remains relatively less explored. By focusing on this aspect, the authors aim to bridge this gap and contribute to the advancement of fracture analysis techniques. The authors believe that integrating bit plane concepts and exploring partial denoising techniques can enhance the accuracy and reliability of fracture classification, thereby improving diagnostic outcomes and patient care in the field of fracture management.

### B. Background

Fracture analysis is a critical task in various scientific and engineering domains, necessitating accurate identification, characterization, and classification of fractures. Over the years, researchers have extensively explored the application of computer vision techniques to enhance fracture analysis and improve diagnostic accuracy.

A research by N et al. [1] focuses on the application of machine learning techniques like Random Forest and neural networks for the classification and detection of bone fractures. Tanzi, Vezzetti, Moreno and Moos [2] presents a comprehensive study on the application of deep learning techniques for bone fracture classification using X-ray images. The authors recognize the significance of automated fracture classification in enhancing diagnostic accuracy, treatment planning, and patient care. Prijs et al. [3] presents the development and external validation of a convolutional neural network (CNN) for first automated delineation (segmentation) of ankle fractures, providing insights into the workings of the models. The paper delves into the internal mechanisms of the CNN, offering in-depth understanding and revealing the complex processes behind its accurate predictions, going beyond its black box nature.

In addition to deep learning, the integration of the bit plane concept has garnered significant attention in medical image research. The utilization of bit plane slicing, a technique that decomposes an image into its binary representation, allows for selective enhancement and noise reduction at different bit levels. This concept has shown promise in enhancing fracture images, reducing noise interference, and extracting informative features. Fraz, Basit, and Barman [4] explores the utilization of morphological bit planes for the extraction of retinal blood vessels. The computation of the blood vascular image are achieved by employing a sequence of morphological operations exclusively on the green channel of an RGB colored retinal image.. The proposed algorithm exhibited a notable average accuracy of 0.9423, surpassing the performance of previous papers in this field. Rizzi and Guaragnella [5] introduce a novel method for skin lesion segmentation on $PH^2$ dataset by employing an image bit-plane multilayer approach. By integrating bit-plane decomposition and the UNET architecture, Tuan et al. [6] have shown that medical professionals can enhance the accuracy and reliability of brain tumor segmentation, which is crucial for diagnosis, treatment planning, and monitoring of brain-related disorders. The research by Chen et al. [7] explores the application of Convolutional Neural Networks (CNN) and bit-plane slicing for breast cancer image classification. The experimental outcomes obtained from analyzing breast cancer image datasets demonstrate that the proposed method exhibits significant enhancements in recognition rates and effectively boosts the overall classification performance when applied to specific bit-planes.

Furthermore, denoising techniques have also been employed in other medical imaging applications. A research by Goyal et al. [8] utilized bit plane slicing as a technique to extract different bit plane slices from the grayscale image. These bit plane slices are subsequently subjected to adaptive bitonic filtering as part of the denoising process. Apart from medical imaging, by applying bit-plane average filtering, Agarwal [9] aims to preserve the important details and edges in the high contrast images while effectively reducing the Gaussian noise interference. The technique takes advantage of the varying levels of noise distribution across different bit planes, enabling adaptive filtering and noise reduction in each plane.

These studies highlight the ongoing efforts in the field of medical imaging, showcasing the integration of various techniques such as deep learning, bit plane slicing, and denoising for fracture analysis and enhanced image quality. These advancements contribute to the development of more precise diagnostic tools, improved treatment planning, and better patient care in the medical domain.

The paper is organized as follows: Section II discusses the datasets used, Section III presents the algorithms employed, Section IV presents the experimental results and accuracy, Section V provides a detailed discussion of the work, and finally, the conclusion is presented in the last section, summarizing the key findings and implications of the research.

## II. MATERIALS

### A. Dataset

This research integrates the FracAtlas dataset, a valuable resource comprising 4,083 X-ray images focused on musculoskeletal bone fractures [10]. With accompanying annotations, this dataset facilitates deep learning tasks, particularly in fracture classification. By leveraging FracAtlas, we train and evaluate our models using diverse annotated fracture images, enabling the development of robust algorithms for fracture analysis. This dataset not only enhances our understanding of fractures but also enables exploration of innovative approaches in medical imaging.

### B. Hardware and Libraries

Python code for model training and inference was developed using Jupyter Notebooks, running on virtual machines provided by Google Colaboratory. Our image processing tasks heavily relied on OpenCV and scikit-image, some powerful open-source computer vision libraries. By leveraging these tools, we ensured efficient execution and effective analysis of our research work.

## III. PROPOSED METHODS

The proposed method for fracture classification integrates multiple components and techniques in medical imaging, including the bit plane concept, denoising filters, partial denoising techniques, and handcrafted feature extraction using Otsu's thresholding. Additionally, deep learning-based extraction with DenseNet121 is employed. Classification algorithms such as Decision Tree and Random Forest are utilized on two types of feature vectors, which includes the combination of 6-bit most significant bits (MSB) with 2-bit denoised least significant bits (LSB) as well as 4 bit from MSB and 4 bit from LSB. The research aims to identify the optimal combination of these approaches for accurate fracture classification.

### A. Bit Plane Slicing

Bit plane slicing technique is used to decompose the intensity values of pixels in an image into their binary representation [11]. Each pixel's intensity value is represented by an 8-bit binary vector, with each bit corresponding to a specific level of significance, ranging from the least significant bit (LSB) to the most significant bit (MSB) as shown in figure 1.

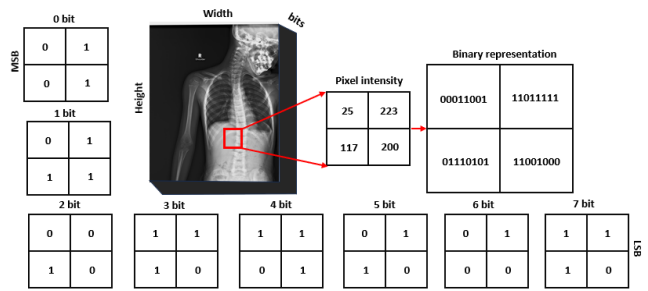

**Figure 1**. Bit Plane Slicing Technique

By segmenting the image into different bit planes based on these binary vectors, valuable information can be extracted. Through the process of binarization, the significant bit planes that contain meaningful information are identified and retained, while the LSB planes are added by utilizing a partial denoising technique. This approach enables the extraction of relevant features, facilitating accurate analysis and interpretation of the image.

### B. Noise Removal

The denoising of the bit planes in this research is accomplished using the Non-local Means Denoising algorithm, implemented through the fastNLMeansDenoising function in the OpenCV library. This algorithm is specifically designed to reduce noise in images by comparing similar patches within the image and estimating noise-free pixel values [12]. By leveraging the capabilities of the fastNLMeansDenoising function, the authors are able to effectively remove noise artifacts from the bit planes while preserving the essential details and structures. Using this function the authors have applied denoising function on the last 2 bit from LSB as proposed partial denoising technique as shown in figure 2.

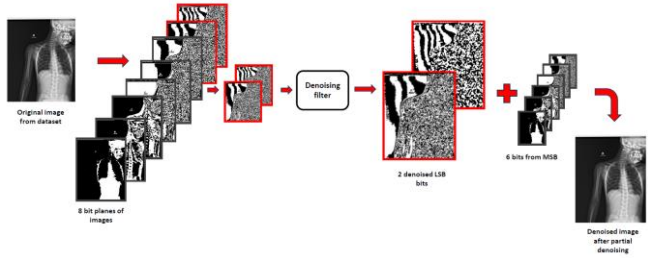

**Figure 2**. Partial Denoising Technique

### C. Feature Extraction

In this study, the authors adopt a comprehensive approach for feature extraction, leveraging the power of deep learning-based feature extraction using the DenseNet121 architecture and the handcrafted feature extraction technique based on Otsu thresholding technique.

*HandCrafted Feature Extraction using OTSU thresholding:*

In handcrafted feature extraction process based on Otsu thresholding [13], the authors performed thresholding on the input image to obtain a binary mask. By multiplying this mask with the original image, the authors isolated the foreground pixels and calculated the mean, variance, and standard deviation as features. Inverting the mask and multiplying it with the original images allowed the authors to extract additional features from the background pixels. In total, six features were extracted (mean, variance, and standard deviation for both foreground and background) from each image, providing a comprehensive representation of the image's characteristics.

*Deep Learning based Feature Extraction using DenseNet:*

In the context of fracture binary classification using the FracAtlas X-ray dataset, DenseNet121 architecture is employed for feature extraction [14]. Leveraging a pre-trained DenseNet121 model pretrained on the ImageNet dataset, the authors fine-tune it specifically for the task of fracture classification. The dense connectivity within DenseNet121 facilitates information flow and enables the extraction of expressive features from the X-ray images. The global average pooling layer condenses the extracted features while preserving important diagnostic information. By utilizing the pre-trained model's learned knowledge, we enhance the accuracy and efficiency of fracture classification on the X-ray dataset.

### D. Classification

Effective classification plays a pivotal role in machine learning, enabling the accurate categorization of data into distinct classes or categories. In this research, the authors focus on leveraging the potential of two robust algorithms, namely decision tree and random forest, for tackling classification tasks. These algorithms have gained significant attention due to their remarkable performance across diverse domains and their ability to handle complex datasets with high-dimensional feature spaces.

*Decision Tree:*

As an integral component of our research methodology, we incorporate the Decision Tree [15] classifier for fracture classification. This classifier utilizes a set of decision rules to partition the data into distinct branches based on various features, enabling the assignment of each data point to a specific fracture class. The authors have utilized default parameter Gini impurity criterion for measuring the quality of a split and the maximum depth of the tree. By leveraging the Decision Tree classifier, the authors' objective is to harness its capability to discern discriminative patterns and achieve precise predictions for fracture classification. To train the Decision Tree model, we utilize the extracted features from the fracture images as input, enabling the model to discern the intricate relationships between these features and their corresponding fracture classes.

*Random Forest:*

In this research, the authors incorporate the Random Forest [16] classifier as a pivotal element for fracture classification. By leveraging the collective knowledge of multiple decision trees, the Random Forest classifier enhances the robustness and accuracy of fracture classification. It leverages the power of feature randomness and bagging techniques to mitigate overfitting and improve generalization performance. The Random Forest classifier utilizes a combination of features extracted from fracture images to train the model, enabling it to capture complex patterns and relationships for accurate classification.

The performance evaluation of both the Decision Tree and Random Forest classifier encompasses crucial metrics such as test accuracy and F1 score, enabling us to comprehensively assess its effectiveness in accurately classifying fractures and contributing to advancements in the field of medical image analysis.

### E. Evaluation Metrics

The fracture classification methodology is evaluated using Signal-to-Noise Ratio (SNR) and Structural Similarity Index Matrix (SSIM) as evaluation metrics. These metrics assess noise reduction, signal fidelity, and preservation of structural information, validating the effectiveness of our methodology in accurately classifying fractures.

*Signal to Noise Ratio (SNR):*

The mean-squared signal-to-noise ratio (SNR) is calculated as the ratio of the signal power to the noise power in the output image. SNR is an important metric for assessing image quality and quantifying the effectiveness of compression algorithms.

The formula of SNR in dB scale is used as an evaluation metric is as follows [17]:

$$SNR_{MS}(dB) = 10\log_{10} \frac{\sum_{x=0}^{M-1}\sum_{y=0}^{N-1} \hat{f}(x,y)^2}{\sum_{x=0}^{M-1}\sum_{y=0}^{N-1}[f(x,y)-\hat{f}(x,y)]^2} \quad (1)$$

In equation 1 $f(x,y)$ indicates the reference image array and $\hat{f}(x,y)$ indicates image array for which SNR is calculated. A higher SNR implies better image quality, as the signal power dominates over the noise power.

*Structural Similarity Index Matrix (SSIM):*
Structural Similarity Index Matrix (SSIM) is a widely used evaluation metric for assessing the structural similarity between two images [18]. It measures the similarity of structural patterns, textures, and details, rather than focusing solely on pixel-level differences. The formula for SSIM is as follows:

$$SSIM(x,y) = \frac{(2\mu_x\mu_y+c_1)(2\sigma_{xy}+c_2)}{(\mu_x^2+\mu_y^2+c_1)(\sigma_x^2+\sigma_y^2+c_2)} \quad (2)$$

In equation 2, $\mu_x$ indicates the average of $\mu_y$, indicates the average of y. Also $\sigma_x^2$ is the variance of x, $\sigma_y^2$ is the variance of y and $\sigma_{xy}$ is the covariance of x and y. Besides, $c_1$ and $c_2$ are two variables to stabilize the division with weak denominator.

A higher SSIM value indicates similarity between the original and compressed images, reflecting better preservation of structural information. By optimizing the SSIM, the authors ensure that the compressed images retain their inherent structural characteristics, resulting in visually appealing representations for fracture classification in medical imaging.

IV. EXPERIMENTATION AND RESULTS

The experimentation process involved the evaluation of performance of different input bit plane combinations for fracture classification. Our focus was on identifying the most effective combination, particularly the integration of 6 MSB and 2 denoised LSB. These combinations, along with the original image, 4 MSB, 4 LSB, and fully denoised images, were systematically evaluated and compared. Figure 3 provides a visual representation of the input bit plane combinations used in this experiment, emphasizing the significance of the 6 MSB and 2 denoised LSB combination. This guided the experimentation process and enabled the authors to identify the optimal combination for achieving accurate and reliable fracture classification results.

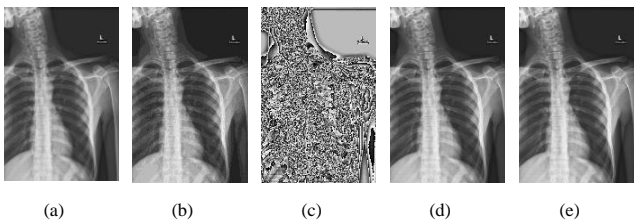

**Figure 3**. (a) Original Image; (b) MSB (4 bit); (c) LSB (4 bit); (d) Partially Denoised Image; (e) Fully Denoised Image

Table 1 showcases the SNR values for the original image, 4 MSB, 4 LSB, the fully denoised image, and a combination of 6 MSB bits and 2 denoised LSB bits. These SNR values offer insights into the effectiveness of each input image set in preserving the signal and minimizing noise.

| Image No | Full Denoised | MSB (4 bit) | LSB (4 bit) | Partial Denoised |
|---|---|---|---|---|
| Image 1 | 35.78 | 20.46 | -20.46 | 22.58 |
| Image2 | 27.28 | 19.8 | -19.8 | 22.86 |
| Image 3 | 37.85 | 17.42 | -17.42 | 20.42 |
| Image 4 | 27.67 | 16.41 | -16.41 | 18.8 |
| Image 5 | 29.56 | 17.27 | -17.27 | 20.32 |

**Table 1**. Signal to Noise Ratio in logarithmic scale (SNR)

Table 2 displays the SSIM values (in percentage) for the different input image sets used in the experimentation process. The SSIM values indicate the structural similarity between the original image and each input set. This validates the effectiveness of this input set in preserving essential fracture features, contributing to accurate fracture classification.

| Image No | Full Denoised | MSB (4 bit) | LSB (4 bit) | Partial Denoised |
|---|---|---|---|---|
| Image 1 | 99.97 | 92.02 | 23.66 | 94.65 |
| Image2 | 99.82 | 96.73 | 15.94 | 96.87 |
| Image 3 | 99.98 | 83.03 | 26.78 | 90.58 |
| Image 4 | 99.85 | 87.84 | 33.88 | 95.11 |
| Image 5 | 99.92 | 92.02 | 25.06 | 93.98 |

**Table 2**. Structural Similarity Index Matrix (SSIM) in percentage

Table 3 displays the classification accuracy and F1 score for features extracted using the Otsu thresholding method, with the Decision Tree and Random Forest classifiers.

| Classifiers | Bit plane Amalgamation | Testing Accuracy | F1 Score |
|---|---|---|---|
| Decision Tree | Original Image | 0.5789 | 0.5695 |
| | MSB (4 bit) | 0.5548 | 0.529 |
| | LSB (4 bit) | 0.5373 | 0.5012 |
| | Full Denoised | 0.5592 | 0.5271 |
| | Partial Denoised | 0.6009 | 0.5864 |
| Random Forest | Original Image | 0.6425 | 0.627 |
| | MSB (4 bit) | 0.603 | 0.5781 |
| | LSB (4 bit) | 0.5943 | 0.5363 |
| | Full Denoised | 0.6272 | 0.6028 |
| | Partial Denoised | 0.6316 | 0.6182 |

**Table 3**. Classifier accuracy based on features extracted by Handcrafted Method (Otsu's thresholding technique)

Table 4 presents the classification accuracy and F1 score obtained by utilizing features extracted using the DenseNet121 architecture.

| Classifiers | Bit plane Amalgamation | Testing Accuracy | F1 Score |
|---|---|---|---|
| Decision Tree | Original Image | 0.7807 | 0.7738 |
| | MSB (4 bit) | 0.8816 | 0.8789 |
| | LSB (4 bit) | 0.7412 | 0.7423 |
| | Full Denoised | 0.8597 | 0.8552 |

|  | Partial Denoised | 0.829 | 0.8219 |
|---|---|---|---|
| Random Forest | Original Image | 0.9364 | 0.9339 |
|  | MSB (4 bit) | 0.9605 | 0.9593 |
|  | LSB (4 bit) | 0.7895 | 0.7876 |
|  | Full Denoised | 0.9364 | 0.9339 |
|  | Partial Denoised | 0.9561 | 0.9539 |

**Table 4**. Classifier accuracy based on deep learning based feature extraction (DenseNet121)

## V. DISCUSSIONS

In this section, the authors investigate noise reduction techniques and classification results, providing insights into the effectiveness of our methodology for accurate fracture classification in medical imaging analysis.

### A. Noise Reduction Analysis

The SNR results in dB scale for table 1 demonstrates that the full denoised bit plane combination consistently achieves the highest SNR values, indicating effective noise reduction and improved signal quality. Additionally, the combination of 6 bit MSB and 2 bit denoised LSB combination outperforms the MSB (4 bit) and LSB (4 bit) combinations, suggesting its potential for higher classification accuracy in fracture analysis.

Similarly, the SSIM values from table 2 show that the last 2 bit denoised and 6 bit MSB combination consistently outperforms the MSB (4 bit) and LSB (4 bit) combinations in terms of structural similarity. This highlights the benefit of selectively denoising the least significant bits while retaining the most significant bits for improved image quality.

To understand why the full denoised image may have better SNR than the partial denoised image, it's important to consider the significance of the MSB bits. The MSB bits contain the most critical information in an image. By not touching these bits in the partial denoised image, the most significant information remains intact and is not negatively impacted by the denoising process.

In contrast, the full denoised image applies noise reduction techniques to all the bits in the image, including the MSB. While this may result in increase of SNR value but it can result in minor loss of significant information from MSB bits.

Overall, these findings highlight the trade-off between noise reduction and feature preservation. While the full denoised combination achieves high SNR values, it may lead to the loss of significant features present in the MSB. The 6 MSB and 2 denoised LSB combination offers a promising alternative by balancing noise reduction and feature preservation, leading to improved classification accuracy.

### B. Classification Result Analysis

The results presented in Table 3 and Table 4 shed light on the performance of different feature extraction techniques and classifiers in the context of OTSU-based and DenseNet-based denoising methods.

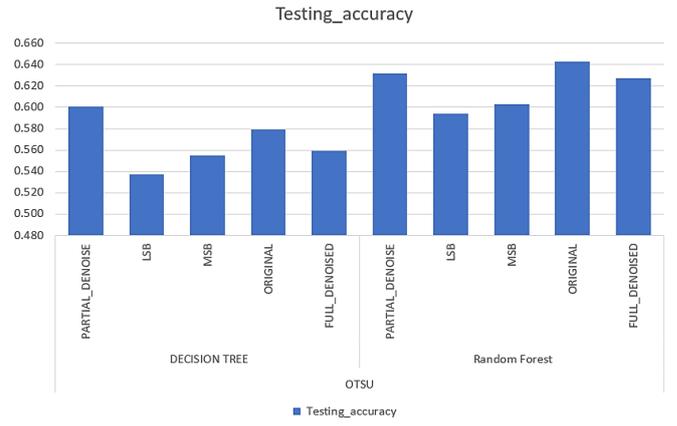

**Figure 4**. Classifier accuracy based on features extracted by Handcrafted Method (Otsu's thresholding technique)

Upon examining the OTSU-based denoising techniques from Figure 4 and Table 3 it is evident that the feature extraction from partially denoised images coupled with the Decision Tree classifier achieves a testing accuracy of 60.08% and an F1 score of 0.586. Comparatively, utilizing the feature extracted from fully denoised images with the same classifier yields a slightly lower testing accuracy of 55.92% and an F1 score of 0.527.

Now in case of the Random Forest classifier, the results are consistent with the trend observed in the Decision Tree classifier. The partial denoised image achieve a higher testing accuracy of 63% and an F1 score of 0.61. Similarly, the full denoised image yields a slightly lower testing accuracy of 62% and an F1 score of 0.60. These findings suggest that incorporating denoising techniques can enhance the classification accuracy, and both partial and full denoised images offer improvements over using the original image alone.

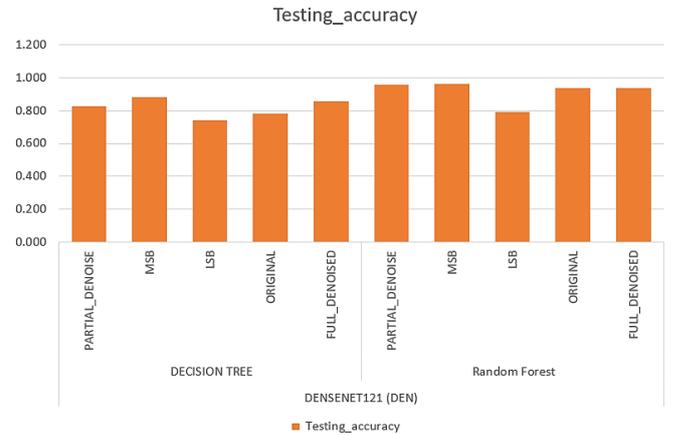

**Figure 5**. Classifier accuracy based on deep learning based feature extraction (DenseNet121)

When considering DenseNet121 as the feature extractor, analyzing the results from Table 4 and Figure 5, it is evident that utilizing the MSB and fully denoised bit planes achieve higher testing accuracies compared to other bit planes in case of Decision Tree Classifier. Specifically, the MSB bit plane achieves a testing accuracy of 88.16%. These results demonstrate the effectiveness of retaining the most significant bits (MSB) in enhancing the classification accuracy.

Similarly, when considering the Random Forest classifier, the MSB and the partially denoised bit planes outperform the other bit planes in terms of testing accuracy. The MSB bit plane achieves a testing accuracy of 96.05%, while the partially denoised bit plane achieves a testing accuracy of 95.61%. These findings further support the superiority of these bit planes in achieving higher classification accuracy.

the superiority of DenseNet121 over OTSU-based denoising techniques can be attributed to its architecture's inherent capacity for image analysis, its dense connectivity facilitating feature learning, and the synergistic combination with decision tree and random forest classifiers. This analysis emphasizes the significance of choosing appropriate feature extraction techniques and classifiers to achieve optimal fracture classification accuracy in medical imaging analysis.

Therefore, the comprehensive analysis of the classification results highlights the potential benefits of integrating denoising techniques and emphasizes the criticality of the feature extraction and classifier selection process. It is crucial to thoroughly assess different combinations to attain optimal accuracy in fracture classification for medical imaging analysis, taking into account the intricacies of denoising, feature extraction, and classification dynamics.

## VI. CONCLUSION

In conclusion, this study presents a comprehensive investigation into fracture classification in medical imaging analysis, with a focus on denoising techniques, feature extraction, classifier selection, and the utilization of bit planes.

The concept of bit planes, as demonstrated in this research, offers a promising avenue for future exploration. Researchers can further investigate the potential of selectively utilizing bit planes to extract and retain the most significant information while reducing noise and enhancing classification accuracy. This can involve exploring different combinations of bit planes, evaluating their impact on feature representation, and assessing their effectiveness across various fracture types and imaging modalities.

Additionally, future research can delve into the development of more advanced algorithms and techniques for bit plane manipulation, such as adaptive thresholding or data-driven approaches, to optimize the extraction of informative features from different levels of significance.

By integrating the concept of bit planes and denoising techniques into future studies, researchers can continue to refine and improve fracture classification algorithms, leading to more accurate diagnoses, better treatment planning, and enhanced patient care in the field of medical imaging analysis.

## VII. ACKNOWLEDGEMENT